\documentclass[letterpaper]{article} 
\usepackage{aaai23}  
\usepackage{times}  
\usepackage{helvet}  
\usepackage{courier}  
\usepackage[hyphens]{url}  
\usepackage{graphicx} 
\usepackage{natbib}  
\usepackage{caption} 
\usepackage{algorithm}
\usepackage{algorithmic}

\usepackage{multirow}
\usepackage{booktabs}
\usepackage{bbding}
\usepackage{arydshln}
\usepackage{amsmath}
\usepackage{newfloat}
\usepackage{listings}
\DeclareCaptionStyle{ruled}{labelfont=normalfont,labelsep=colon,strut=off} 
\floatstyle{ruled}
\newfloat{listing}{tb}{lst}{}
\floatname{listing}{Listing}
\nocopyright
\title{SeDR: Segment Representation Learning for Long Documents Dense Retrieval}
\author{
    Junying Chen\textsuperscript{\rm 1},
    Qingcai Chen\textsuperscript{\rm 1,2}\thanks{Corresponding author.},
    Dongfang Li\textsuperscript{\rm 1},
    Yutao Huang\textsuperscript{\rm 1}
}
\affiliations{
    \textsuperscript{\rm 1} Harbin Institute of Technology, Shenzhen\\
    \textsuperscript{\rm 2} Peng Cheng Laboratory\\
    \{junying.chen.cs, crazyofapple\}@gmail.com, qingcai.chen@hit.edu.cn, yutaohuang7777@163.com
}

\usepackage{bibentry}

\begin{document}

\maketitle

\begin{abstract}
Recently, Dense Retrieval (DR) has become a promising solution to document retrieval, where document representations are used to perform an effective and efficient semantic search. However, DR remains challenging on long documents, due to the quadratic complexity of its Transformer-based encoder and the finite capacity of a low-dimension embedding. Current DR models use suboptimal strategies such as truncating or splitting-and-pooling to long documents leading to poor utilization of whole document information. In this work, to tackle this problem, we propose \textit{Se}gment representation learning for long documents \textit{D}ense \textit{R}etrieval (SeDR). In SeDR, Segment-Interaction Transformer is proposed to encode long documents into document-aware segment representations, while it holds the complexity of splitting-and-pooling and outperforms another segment-interaction patterns on DR. Since GPU memory requirements for long document encoding cause insufficient negatives for DR training, Late-Cache Negative is further proposed to provide additional cache negatives for optimizing representation learning. Experiments on MS MARCO and TREC-DL datasets show that SeDR achieves superior performance among DR models, and confirm the effectiveness of SeDR on long document retrieval.

\end{abstract}

\section{Introduction}
Document retrieval is a crucial component in information retrieval (IR) and benefits for various IR-related tasks, e.g. web search~\cite{xiao2022progressively} and question answering~\cite{inbatch1}. Despite its significance, document retrieval is a challenging problem that requests to rapidly retrieve top documents from scale documents collection. Traditional bag-of-words (BoW) models such as BM25~\cite{bm25} conduct it by term exact matching and therefore suffer from the term mismatch problem~\cite{star}. Recently, with the progress of pre-trained language models like BERT/RoBERTa~\cite{devlin2018bert,liu2019roberta}, dense retrieval (DR) has shown promising results in document retrieval that significantly outperforms BoW models, benefiting from their powerful semantic representative ability. For a typical DR model, query and document are encoded into single low-dimension dense embedding. Then a dot-product or cosine function is applied to them for computing their relevance score. With the offline-constructed index of document embedding, dense retrieval can fulfill efficient semantic search in milliseconds online.

\begin{figure}[t]
  \centering
  \includegraphics[width=0.95\columnwidth]{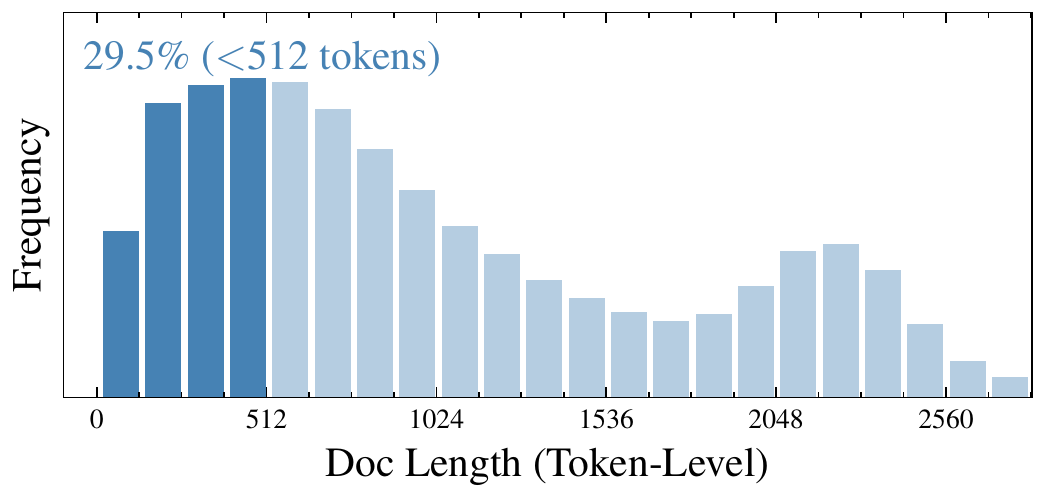}
  \caption{\label{plt4} Document Length Distribution of MS MARCO}
\end{figure}

However, current DR models are unfit for long documents due to the following problems: \textbf{(1) Length Limitation.} As Transformer-based pre-trained models are used in DR models, the length limitation in Transformer is inevitable. Since the memory and computational requirements scale quadratically $O(n^2)$ with text length $n$, long documents can't be input directly. For instance, as shown in Figure~\ref{plt4}, only 29.5\% documents are less than 512 tokens, which is a maximum length of BERT-base. To tackle this issue, prior work adopts bandit solutions such as truncating~\cite{inbatch1,star} or splitting-and-pooling~\cite{ance}, both of which cause information loss. Truncating the long documents to meet input length loses the information from unused segments. Splitting the documents into text segments and max-pooling the segment scores can obviate unused segments, but violate the consistency and relevance of segments in the document by encoding each segment independently. \textbf{(2) Finite Representation Capacity.} According to the analysis in \citet{mebert} and \citet{Multi-View}, the finite capacity of a single low-dimension embedding is awkward to represent a document well, especially for the long document that covers more diverse information. To enhance the representation capacity, recent studies resort to multiple representations for document~\cite{Multi-View}. Nevertheless, the effectiveness of multiple representation comes with extra computation and memory demand in searching. Due to the dispersed length of realistic documents as shown in Figure~\ref{plt4}, these methods are not reasonable to use fixed-number vectors to represent both short and long documents. Moreover, our empirical experiment reveals that multiple representations tend to collapse into the same one, which constrains the capacity of multiple representations~\cite{Multi-View}. \textbf{(3) Memory Bottleneck.} Recent studies make progress on negative sampling strategies for DR training. ~\citet{morenegative} found that more contrastive negatives often lead to better performance. Hence, the prior works introduce in-batch negative sampling~\cite{inbatch1,inbatch2}, which utilizes other document representations in mini-batch as negatives without extra encoding cost. However, in-batch negative is limited to the batch size that GPU memory can sustain. Along with text length increasing, long document encoding consumes more GPU memories, thus leading to insufficient negatives for representation learning.

To address these problems, we propose \textit{Se}gment representation learning for long documents \textit{D}ense \textit{R}etrieval (SeDR), which split long document to segments and transfer them into document-aware segment embeddings for DR. Since segment embedding represents a fixed-length segment so that the number of embeddings depends on document length, segment representations can exert their capacity to represent non-redundant information. Meanwhile, the relevance matching between queries and documents usually takes place at text segments of document~\cite{QDS}, which illustrates that segments of document are distinct for searching. Therefore, segment representation learning can be a more effective and efficient paradigm for DR. 

Despite these merits, segment representations suffer from the independence bias on splitting-and-pooling as stated before. To solve this issue, we propose Segment-Interaction Transformer. It upgrades the Transformer with a simple modification, enabling segments from the same document to interact with each other in self-attention mechanism for generating document-aware segment representations with less information loss from document splitting. Moreover, Segment-Interaction Transformer preserves the capacity of pre-trained model like BERT/RoBERTa with negligible extra parameters, and hold the complexity $O(n_s^2)$, where $n_s$ denotes the segment length, instead of original $O(n^2)$. Extensive experiments show that Segment-Interaction Transformer outperforms other segment-interaction patterns, e.g. Longformer\cite{beltagy2020longformer} and global-attention~\cite{globalattention1}, on DR with the smallest parameters and highest efficiency.

There is also a problem with memory limitation that result in insufficient negatives for training on long documents. To tackle it, we present Late-Cache Negative, which stores late encoding representations as cache negatives and introduces them to improve subsequent training in a free-cost way. Since the cache negatives don’t require gradient, the number of cache negatives is not limited to GPU memory so that it can provide extra negatives for training. Our experiments show that the extra cache negatives are necessary to improve the performance for long document training. The code is available at \url{https://github.com/jymChen/SeDR}.

Our main contributions can be summarized as follows:
\begin{itemize}  
    \item To the best of our knowledge, this is the first study to address the issues of long document on DR. And we propose a Segment-Interaction Transformer to encode long documents efficiently into document-aware segment representations that outperforms other segment-interaction patterns on DR.
    \item We propose Late-Cache Negative to break down the constraint of GPU memory and provide additional negative instances, which is verified to be necessary for long documents DR in the ablation study. We also further investigate the effectiveness of different negative settings.
    \item Our model SeDR achieves superior retrieval performances among DR models on the MSMARCO-Document dataset as well as TREC-DL 19' and 20'.
\end{itemize}

\section{Related Work}
\paragraph{Dense Retrieval} 
Dense retrieval has been significantly promoted by Negative sampling. Recent studies prove that sampling "hard" negatives for contrast learning benefit for DR. \citet{bm25neg,inbatch1} sample hard negatives from BM25 top documents. In \citet{qu2020rocketqa,ance}, they sampled harder negative from the top documents retrieved by the warm-up DR model. \citet{star} is proposed to optimize query encoder with dynamic hard negative sampled from ANN-searched documents. Moreover, the in-batch negative sampling is introduced to augment negatives~\cite{in-batch3,in-batch4}. Later on, \citet{qu2020rocketqa,xiao2021matching} adopt the cross-device negative sampling to promote in-batch negatives by using the negatives from other distributed devices, which is still confined to device memory. Besides, there are some studies that try to employ multiple representations to improve the representation ability of documents. \citet{mebert} utilizes the first k document token embeddings as the document representation. \citet{Multi-View} is proposed to use multiple viewer tokens replacing single \emph{[CLS]} to generate document vectors. \citet{colbert} stores each token embeddings for later interaction, which is unfit to document retrieval due to the tokens scale of long documents. 

\paragraph{Long Document Solution} For long documents, most previous DR models take the first 512 tokens as input~\cite{star,qu2020rocketqa,inbatch1}. \citet{ance} extra use the MaxP operation~\cite{maxp} to split the document into segments and max-pooled the segment scores. Despite the convenience of these bandit solutions, they cause significant information loss of long documents~\cite{QDS}. There are several works proposed to promote the long document ranking on long documents~\cite{TKL,QDS,zhou2022socialformer,fu2022leveraging}. However, these reranking models are impractical to rank all documents in corpus to do retrieval that would incur severe calculation delay. Another idea to solve the long document problem is to use sparse attention patterns~\cite{beltagy2020longformer,zhou2022socialformer,kitaev2020reformer}, which can avoid quadratic complexity in self-attention mechanism. One of the most successful sparse attention is the sliding window attention~\cite{hofstatter2020local,qiu2019blockwise,beltagy2020longformer}, which allows each token to attend to its surrounding tokens. In \citet{globalattention1,bigbird,gupta2020gmat}, global attention is proposed to fit the specific tasks. Among long-document Transformers, Longformer~\cite{beltagy2020longformer} is a prevalent model that integrates the sliding window attention and the global attention.

\section{Preliminaries}
\subsection{Task Definition}
Given a query $q$ and a massive document collection $D$, a document retrieval task aims to find the positive document $d^+ \in D$, or provide high-quality candidates for further reranking tasks where it served as a first-stage retrieval.  

\subsection{Dense Retrieval Architecture}
By precomputing the index of document representations, DR has been proved effective and efficient in document searching. We start with introducing a common dense retrieval architecture for document retrieval. A typical DR model encodes query $q$ and document $d$ separately into dense representations using a query encoder $E_Q$ and a document encoder $E_D$ respectively. In this process, these encoders widely adopt \emph{[CLS]} representations from pre-trained language models such as BERT/RoBERTa. Then, a similarity function $sim(\cdot)$, usually inner product, is used to perform efficient retrieval by predicting the similarity score $f(q, d)$ of query $q$ and document $d$:
\begin{equation}
\label{sim}
    f(q,d)=sim(E_{Q}(q),E_{D}(d))
\end{equation}
Supervised by the training set of the target retrieval task, query and document encoders are trained with a contrastive-learning loss. In our models, we use InfoNCE~\cite{infonce} that computes the loss for a given instance $q$ as:
\begin{equation}
\label{loss1}
    \mathcal{L}(q,d^+)=-log{\frac{e^{f(q,d^+)}}{e^{f(q,d+)}+\sum\limits_{d\in D^-_q}{e^{f(q,d)}}}}
\end{equation}
where $D_q^-$ is the negative document set that contrasts with the positive one $d^+$ for representation learning. In practice, $|D_q^-|<<|D|$ since the cost for computation of all negative documents is prohibitive. To improve the performance, hard negative and in-batch negative are introduced to sample negatives. Hard negative strategy is used to sample top-K documents of $q$ as hard negative $d^-$:
\begin{equation}
\label{sample}
    d^-=\mathrm{Sample}(\mathrm{Top}{-}K(q))
\end{equation}
Thus we call K as the hardness of hard negative to distinguish learning. In-batch negative sampling leverages the positive document and the negative document from other queries in the same mini-batch, served as random negatives in a cost-free way, where each training instance is used for the augmentation of other instances' negative samples. With these two negative sampling strategies, the negative sample set $D_q^-$ become:
\begin{equation}
\label{negatives}
    D^-_q={d^-}\cup\{d^+,d^-\}_{q^\prime \in B\land q^\prime \neq q}
\end{equation}
where B denotes the query set in a mini-batch. 

\section{SeDR}
SeDR is built on the above DR architecture and proposed to enhance it on long documents. It comprises of Segment-Interaction Transformer and Late-Cache Negative, both essential to improve retrieval performance for long documents.

\subsection{Segment-Interaction Transformer}
\begin{figure*}[ht]
  \centering
  \includegraphics[width=0.9\textwidth]{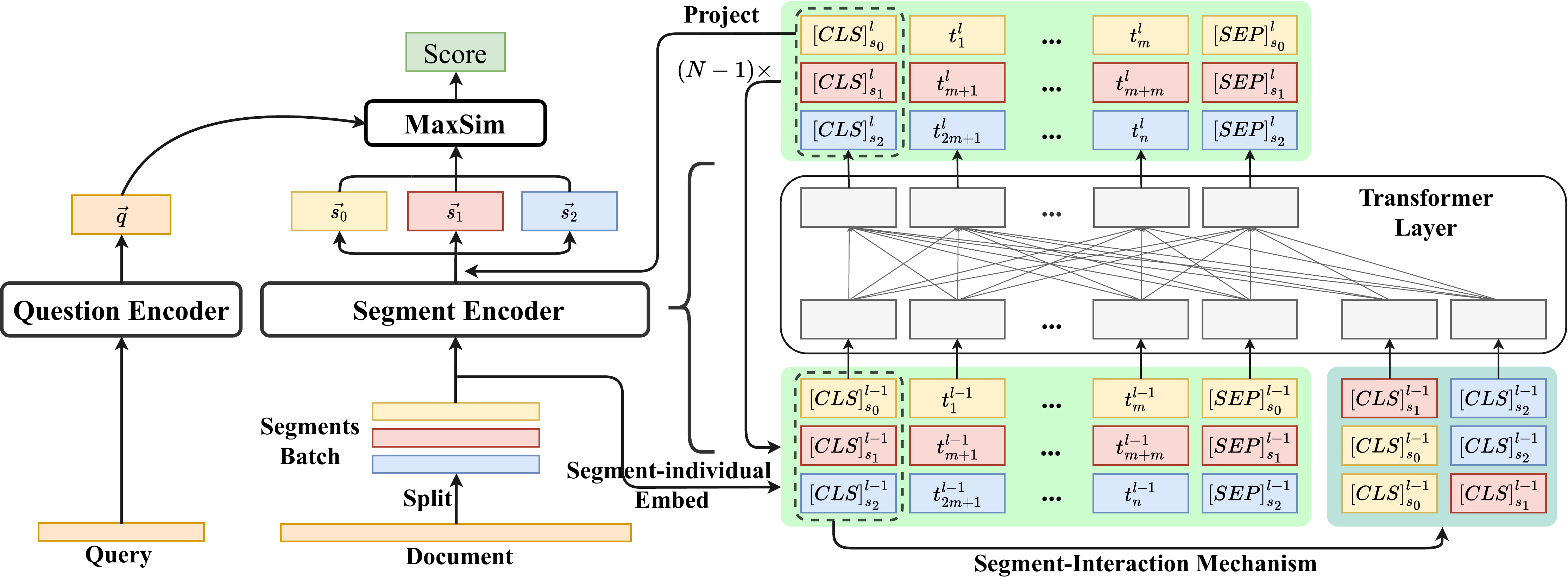}
  \caption{\label{overview}Illustration of segment representation learning with Segment-Interaction Transformer.}
\end{figure*}
Given a n-tokens document $d=[t_1,t_2,...,t_n]$, it firstly splits the document to $m$-tokens segments batch $[s_1,...,s_k]$, where the segment number $k=\left \lceil \frac{n}{m} \right \rceil$ and the last segment $s_k$ will be padded to $m$ tokens. Then each segment will be concatenated with the special tokens according to the pre-trained model as:
\begin{equation}
\label{segment}
    s_{i}=[CLS,t_{im+1},...,t_{im+m},SEP]
\end{equation}
where $s_i$ indicates the $i$-th segment in the segment batch. Unlike the splitting-and-pooling method, the segment batch will be input into the Segment-Interaction Transformer to obtain the document-aware segment representations.

\paragraph{Segment-individual Embedding} 
After the input tokens are converted to token embedding, position embedding is added to them to signal the position so that the model can perceive the order of sequence. Similarly, we introduce the segment embedding to indicate the segment order in documents. Specifically, the input representation $H_{s_i}^0$ of segment $s_i$ is computed by summing token embedding, position embedding and the $i$-th segment embedding.

\paragraph{Segment-Interaction mechanism} 
Following previous DR models, we use BERT/RoBERTa architecture that comprises of multiple stacked bidirectional transformer layers. The key point of the transformer layer is self-attention mechanism~\cite{vaswani2017attention} that calculates the $l$-th layer output $H^l$ taking the input $H^{l-1}$ from the previous layer:
\begin{equation}
\label{attention1}
\begin{gathered}
    Q^T;K^T;V^T = (W^q;W^k;W^v)\cdot H^{l-1}_{s_i} \\
    H^l_{s_i} = \mathrm{softmax}(\frac{Q K^T}{\sqrt{d_k}})\cdot V^T
\end{gathered}
\end{equation} 
where $H^l_{s_i}$ denotes the $l$-th layer output of segment $s_i$ and $W^q;W^k;W^v$ are three linear projections of query, key and value. In the self-attention layer, each token can attend to other tokens in the same segment, but fail to attend to other segment tokens. To fulfill segments interaction, we propose a Segment-Interaction mechanism that is simple but effective. As shown in Figure~\ref{overview}, the hidden representation $H_{s_i}^{l-1}$ is integrated with the \emph{[CLS]} representations from other segments to produce a new $H_{s_i}^{l}$ in each layer as:
\begin{equation}
\label{attention2}
\begin{gathered}
    H_{s_i}^{l-1}=[h_{CLS}^{s_i,l-1},h_{t_{im+1}}^{s_i,l-1},...,h_{t_{im+m}}^{s_i,l-1},h_{SEP}^{s_i,l-1}] \\
    Q^T = W^q\cdot H^{l-1}_{s_i} \\
    \dot{K}^T;\dot{V}^T = (W^k;W^v)\cdot [H^{l-1}_{s_i}\circ [h_{CLS}^{s_j,l-1}]_{j\neq i}] \\
    H^l_{s_i} = \mathrm{softmax}(\frac{Q\dot K^T}{\sqrt{d_k}})\cdot \dot{V}^T
\end{gathered}
\end{equation}

where $\circ$ is the concatenation operation, $h_{CLS}^{s_i,l-1}$ denotes the \emph{[CLS]} representation of segment $s_i$ on ($l-1$)-th layer output and $[h_{CLS}^{s_j,l-1}]_{j\neq i}$ denotes the list of \emph{[CLS]} representations on other segments built automatically. In this way, each \emph{[CLS]} token can attend to its segment tokens as well as other segment \emph{[CLS]} tokens, while segment tokens can also attend to other segment \emph{[CLS]} tokens. Hence, different segments interact across segment batch using the bond of \emph{[CLS]} representations, where segment tokens can capture other segments information and focus on its own segment content. Meanwhile, \emph{[CLS]} are controlled to attend to its segment content and interact with other segment representation in 
every Transformer layers to perceive global information of the document. More importantly, this upgrade is very cost effective in practice. Then we use a linear layer on the final output \emph{[CLS]} representations to gain the segment representations:
\begin{equation}
\label{segInteraction}
    [\vec{s_1},...,\vec{s_k}]=[h_{CLS}^{s_1},...,h_{CLS}^{s_k}]\times W+b
\end{equation} 
where $\vec{s_i}$ denotes the $i$-th segment representation and $h_{CLS}$ denotes the final output of \emph{[CLS]} representations.

\paragraph{Pooling} 
To compute the score of query $q$ and document $d$, a max-pooling is applied to multiple segment representations of $d$ to calculate the final score as:
\begin{equation}
\label{score}
    f(q,d)=\max\limits_{i}\left \{ \mathrm{sim}(E_Q(q), \vec{s_i} )\right \}
\end{equation} 
Thus, the top-score documents can be retrieved with one fast approximate nearest neighbor (ANN) search operation during inference, using the index of segment embedding that is built offline. Therefore, segment representations can remain the advantage of fast search in dense retrieval.

\begin{figure}[t]
  \centering
  \includegraphics[width=0.83\columnwidth]{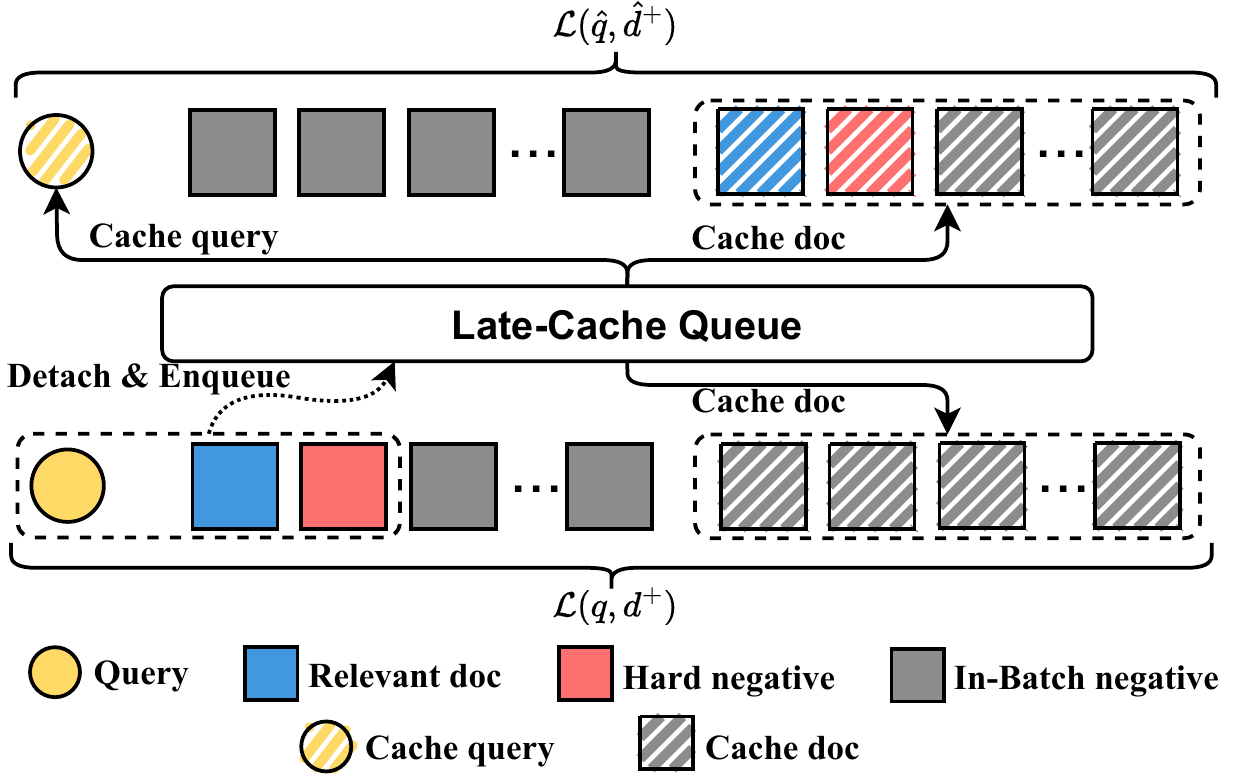}
  \caption{\label{cacheneg}Schematic of Late-Cache Negative.  }
\end{figure}
\subsection{Late-Cache Negative}
A longer document demands more memory when encoding, resulting in a smaller batch-size $B$, as well as fewer negatives $|D_q^-|$ for $\mathcal{L}(q,d^+)$ on DR training. Despite the efficiency of In-batch negative, it's not flexible that the number of negatives is limited to the batch-size setting. Thus, we propose Late-Cache Negative that provides more negatives in a cost-free way without the limitation of GPU memory or batch-size setting. 

The point of the idea is to keep a late-cache queue $Q$, that stores the late-computed representations of documents and queries. For a training instance $\{q,d^+,d^-\}$, where $d^+$ and $d^-$ indicate the positive document and hard negative of $q$, the embeddings of $\{q,d^+,d^-\}$ will detach the training gradient and enqueue to the queue $Q$ after the training step. Because the cache embeddings don't require gradients, the $Q$ can store a large number of embeddings to provide enough cache negatives. And the cache-size $C$ is a hyperparameter that can flexibly control the number of cache negatives used in training. As illustrated in Figure~\ref{cacheneg}, the cache negatives in $Q$ are used in two ways. Firstly, the cache negatives provide the cache documents to $D^-_q$ that:
\begin{equation}
\label{cache1}
    D^-_q={d^-}\cup\{d^+,d^-\}_{q^\prime \in B\land q^\prime \neq q}\cup\{\hat d^+,\hat d^-\}_{\hat q \in Q}
\end{equation}
Consequently, the extra negatives of $Q$ can supply extra random negatives to replenish the insufficient negatives set for long document representation learning. On the other hand, the cache query $\hat q$ in $Q$ is used to further constrict the document representations learning as:
\begin{equation}
\label{cache2}
\begin{gathered}
    \mathcal{L}(\hat q, \hat{d}^+) = -log{\frac{e^{f(\hat q, \hat{d}^+)}}{e^{f(\hat q, \hat{d}^+)}+\sum\limits_{d\in D^-_{\hat{q}}}e^{f(\hat{q},d)}}} \\
    D^-_{\hat{q}}={\hat{d}^-}\cup\{d^+,d^-\}_{q \in B}\cup\{\hat{d}^+,\hat{d}^-\}_{\hat{q} \prime \in Q \land \hat  q^\prime \neq \hat q}
\end{gathered}
\end{equation}
where the document embeddings are trained to keep away from previous document embeddings and query embeddings and adapt the distribution in the dense space. In the end, the final training loss of a training batch is computed as:
\begin{equation}
\label{cache4}
    \mathcal{L}_{\mathrm{batch}}= \sum_{q\in B}{\mathcal{L}(q, d^+)}+\sum_{\hat q\in Q}{\mathcal{L}(\hat q,\hat d^+)}
\end{equation}

\section{Experiments}
\subsection{Datasets}
We evaluate SeDR on the TREC Deep Learning Track benchmarks~\cite{DL2019,DL2020} for document retrieval. Specifically, we use its MSMARCO-Document corpus~\cite{msmarco} that contains 3,213,835 documents, 367,013 training queries and 5,193 Dev queries (MARCO Dev), where each query has one relevant-judged document. In addition, we utilize two differently-constructed test datas: 2019 Deep Learning track~\cite{DL2019} (TREC 19' DL) with 43 queries and 2020 Deep Learning track~\cite{DL2020} (TREC 20' DL) with 45 queries, where each query involves multiple relevant documents. Following \citet{ance}, we truncate the documents to a maximum of 2048 tokens and report the official metrics and Recall@100 from the full-corpus retrieval results.

\subsection{Baselines}
\paragraph{Retrieval Models} 
BM25~\cite{bm25} and docT5query~\cite{nogueira2019doc2query} are used as two BoW baselines, where we use Anserini implementation~\cite{yang2018anserini}. For DR models, we adopt three DR baselines with different sampling strategies: Rand Neg~\cite{randneg}, BM25 Neg~\cite{bm25neg} and In-batch Neg~\cite{inbatch1,inbatch2}. Other competitive DR models are ANCE~\cite{ance}, which iteratively samples hard negatives of the model, and STAR~\cite{star} that improve ANCE by introducing in-batch negative. To fit long documents, we apply MaxP~\cite{maxp} operation to upgrade ANCE and STAR, where the document is split into 512-token segments and the scores are max-pooled. To evaluate the multiple representations, STAR-multi is introduced to train STAR to generate multiple vectors using viewer tokens in \citet{Multi-View}, where the vector number is set to 4. We also compare the DR models under ADORE~\cite{star} mechanism, which is used to further optimize the query encoder to better fit on the document index.

\paragraph{Segment-Interaction Baselines} 
To evaluate our Segment-Interaction Transformer, we change the segment-interaction pattern of SeDR and obtain 4 comparable models. 1) SeDR-MaxP, where MaxP operation is adopted on SeDR meaning that all the segment representations are independent. 2) SeDR-Transformer-Head, where the segment representations can interact via a single extra Transformer layer on top of SeDR-MaxP. 3) SeDR-Global-Attention, where the global attention is utilized to enable the each segment \emph{[CLS]} token attend to whole document tokens. 4) SeDR-Longformer, where Longformer is used to encode the long document and generate the segment representation by inserting \emph{[CLS]} tokens in document.

\subsection{Implementation Details}
All DR models are based on RoBERTa-base model. For document splitting, we set the segment length as 512 and maximum number of segments as 4. For training settings, we use STAR as warm-up model to initialize SeDR as well as other upgrade models including STAR-multi and STAR(MaxP) to further train it on long document input for fair comparisons. Thus, we use STAR to generate top candidates as the static hard negatives. Following \citet{star}, we sample one hard negative from top-200 documents. When using Late-Cache Negative, we set the cache size as 50 and sample hard negative from top-100 documents. We use Lamb optimizer~\cite{lamb} with a learning rate 5e-5 for all experiment settings. All model training are conducted on Tesla V100 (32G) GPUs and every model is limited to access to one GPU. Thus, the training batch size is set to 17, a maximum can be took in the GPU memory. For inference, we adopt the Faiss library~\cite{faiss} to perform efficient similarity search.

\begin{table*}[t]
\centering
\begin{tabular}{lccccccccc} 
\toprule
\multirow{2}{*}{}       & \multirow{2}{*}{\begin{tabular}[c]{@{}c@{}}Long \\input\end{tabular}} & \multicolumn{2}{c}{MARCO Dev} & \multicolumn{2}{c}{TREC 19' DL} & \multicolumn{2}{c}{TREC 20' DL} & \multirow{2}{*}{\begin{tabular}[c]{@{}c@{}}index \\size\end{tabular}} & \multirow{2}{*}{\begin{tabular}[c]{@{}c@{}}Latency\\~(ms)\end{tabular}}  \\ 
\cmidrule(l){3-4} \cmidrule(l){5-6} \cmidrule(l){7-8}
                        &                                                                       & MRR            & Recall           & NDCG           & Recall             & NDCG           & Recall             &                                                                       &                                                                          \\ 
\midrule
\multicolumn{10}{l}{\textbf{BoW model }}                                                                                                                                                                                                                                                                                                                           \\
BM25                    & \scriptsize{\Checkmark}                                             & 0.277          & 0.807            & 0.519          & 0.385              & 0.506          & 0.586              & -                                                                     & 87.2                                                                     \\
docT5query              & \scriptsize{\Checkmark}                                             & 0.327          & 0.861            & 0.597          & 0.399              & 0.582          & 0.618              & -                                                                     & 91.5                                                                     \\ 
\midrule
\multicolumn{10}{l}{\textbf{Dense Retrieval}}                                                                                                                                                                                                                                                                                                                      \\
In-Batch Neg            & \scriptsize{\XSolidBrush}                                           & 0.320          & 0.864            & 0.544          & 0.295              & 0.509          & 0.479              & 9.2G                                                                  & 1.3                                                                      \\
Rand Neg                & \scriptsize{\XSolidBrush}                                           & 0.330           & 0.859            & 0.572          & 0.284              & 0.500          & 0.484              & 9.2G                                                                  & 1.3                                                                      \\
BM25 Neg                & \scriptsize{\XSolidBrush}                                           & 0.360          & 0.877            & 0.597          & 0.268              & 0.569              &  0.464                 & 9.2G                                                                  & 1.3                                                                      \\
ANCE            & \scriptsize{\XSolidBrush}                                           & 0.377          & 0.893            & 0.615          & 0.277              & 0.580          & 0.497              & 9.2G                                                                  & 1.3                                                                      \\
STAR                    & \scriptsize{\XSolidBrush}                                           & 0.390          & 0.913            & 0.605          & 0.313              & 0.575          & 0.513              & 9.2G                                                                  & 1.3                                                                      \\
STAR-Multi              & \scriptsize{\XSolidBrush}                                           & 0.404          & 0.913            & 0.616          & 0.309              & 0.596          & 0.508              & 36.8G                                                                 & 4.8                                                                      \\
ANCE(MaxP)              & \scriptsize{\Checkmark}                                             & 0.384          & 0.906            & 0.628          & 0.323              & 0.612          & 0.540              & 21.5G                                                                 & 2.7                                                                      \\
STAR(MaxP)              & \scriptsize{\Checkmark}                                             & 0.394          & 0.909            & 0.615          & 0.302              & 0.586          & 0.510              & 21.5G                                                                 & 2.7                                                                      \\ 
\hdashline
SeDR(Ours)              & \scriptsize{\Checkmark}                                             & \textbf{0.409} & \textbf{0.921}   & \textbf{0.632} & \textbf{0.343}     & 0.607          & 0.527              & 21.5G                                                                 & 2.7                                                                      \\
\hspace{1em} \small{w/o Segment-Interaction} & \scriptsize{\Checkmark}                                             & 0.403          & 0.917            & 0.611          & 0.312              & 0.594          & 0.511              & 21.5G                                                                 & 2.7                                                                      \\
\hspace{1em} \small{w/o Late-Cache Negative} & \scriptsize{\Checkmark}                                             & 0.400          & 0.915            & 0.616          & 0.340              & \textbf{0.617} & \textbf{0.551}     & 21.5G                                                                 & 2.7                                                                      \\ 
\midrule
\multicolumn{10}{l}{\textbf{Dense Retrieval w/ ADORE}}                                                                                                                                                                                                                                                                                                             \\
STAR + ADORE            & \scriptsize{\XSolidBrush}                                           & 0.405          & 0.919            & 0.628          & 0.317              & 0.604          & 0.519              & 9.2G                                                                  & 1.3                                                                      \\
STAR-Multi + ADORE      & \scriptsize{\XSolidBrush}                                           & 0.417          & 0.919            & 0.634          & 0.326              & 0.594          & 0.519              & 36.8G                                                                 & 4.8                                                                      \\
ANCE(MaxP) + ADORE      & \scriptsize{\Checkmark}                                             & 0.396          & 0.915            & 0.627          & 0.315              & 0.612          & 0.542              & 21.5G                                                                 & 2.7                                                                      \\
STAR(MaxP) + ADORE      & \scriptsize{\Checkmark}                                             & 0.414          & 0.921            & 0.631          & 0.319              & 0.577          & 0.515              & 21.5G                                                                 & 2.7                                                                      \\
SeDR + ADORE            & \scriptsize{\Checkmark}                                             & \textbf{0.421} & \textbf{0.933}   & \textbf{0.645} & \textbf{0.353}     & \textbf{0.626} & \textbf{0.549}     & 21.5G                                                                 & 2.7                                                                      \\
\bottomrule
\end{tabular}
\caption{Evaluation of retrieval models on TREC Deep Learning Track for document retrieval. MRR, NDCG and Recall are set to MRR@100, NDCG@10 and Recall@100. Latency indicates average time requirement for one query searching.}
  \label{res1}
\end{table*}

\subsection{Comparison with Retrieval Models}
In this section, we analyse the effectiveness of different document retrieval models. We conduct the experiments on TREC Deep Learning Track for document retrieval and report the results in Table~\ref{res1}. In the results, DR models outperform BoW models by a large margin, except for the Recall@100 metric on TREC DL Doc, which may be caused by unlabeled relevant documents~\cite{ance}. Besides, there is much less retrieval latency on DR models due to their GPU acceleration.

\begin{figure}[t]
  \centering
  \includegraphics[width=0.95\columnwidth]{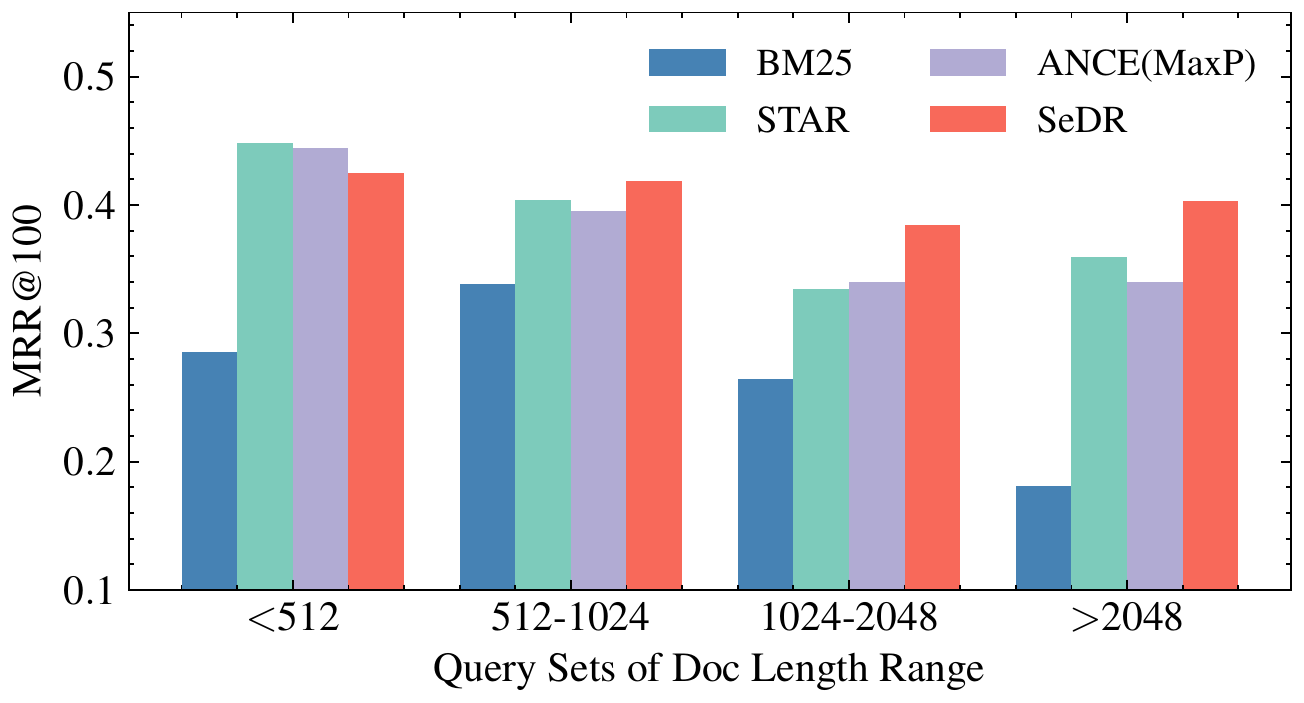}
  \caption{\label{plt2} Performance with query sets of different document length ranges on MARCO Dev.}
\end{figure}

Among DR models, SeDR achieves remarkable performance. Firstly, it significantly surpasses the previous strongest model STAR by a large margin on MARCO Dev and TREC DL, which is consistent with expectation to make full use of long document information. Secondly, for the models taking long document input without truncation, SeDR outperforms ANCE(MaxP) and STAR(MaxP) within the same index size and retrieval latency. Specifically, we further compare the retrieval performance in different document lengths and report the results in Figure~\ref{plt2}. In Figure~\ref{plt2}, SeDR obviously outperforms other models on longer documents, while performing worse than the ANCE(MaxP) and STAR on the documents that are less than 512 tokens. It demonstrates that SeDR can considerably improve the DR performance on the retrieval of longer documents, and  better learn long document representations than conventional MaxP fashion. Thirdly, compared with multi-vector model STAR-Multi, SeDR captures better retrieval results with much less index size. Larger index introduces more retrieval latency as well as more memory occupation. To alleviate the redundancy of multiple vectors in short documents, SeDR uses segment vectors whose number depends on the document length, instead of generating fixed-number vectors for each document. Fourthly, with ADORE, SeDR achieves greater performance and overwhelms other models. Since the ADORE only optimizes the query encoder, the significant improvement on SeDR is due to its document index, i.e. segment representations. It indicates that SeDR can learn better representations for long documents dense retrieval.

Besides, we conduct an ablation study with SeDR to evaluate the contribution of its two components: Segment-Interaction Transformer and Late-Cache Negative. As shown in Table~\ref{res1}, we can see that without Segment-Interaction Transformer or Late-Cache Negative, SeDR performs worse than the original model but still outperforms STAR(MaxP). It verifies that Segment-Interaction Transformer and Late-Cache Negative both contribute to long document DR. Furthermore, it can be observed that SeDR without Late-Cache Negative performs best in TREC 20' DL, which may be caused by greater generalization without Late-Cache Negative that would introduce more negatives for training.

\begin{table*}[t]
\centering
\begin{tabular}{lcccccccccc} 
\toprule
\multirow{2}{*}{}     & \multicolumn{2}{c}{MARCO Dev~}  & \multicolumn{2}{c}{TREC 19' DL} & \multicolumn{2}{c}{TREC 20' DL} & \multicolumn{2}{c}{Time(hour)}   & \multirow{2}{*}{\#Param} & \multirow{2}{*}{\begin{tabular}[c]{@{}c@{}}Latency\\~(ms)\end{tabular}}  \\ 
\cmidrule(lr){2-3}\cmidrule(lr){4-5}\cmidrule(lr){6-7}\cmidrule(lr){8-9}
                      & MRR            & Recall         & NDCG           & Recall         & NDCG           & Recall         & \small{Training} &   \small{Indexing}                         &                                                                          \\ 
\midrule
SeDR-MaxP             & 0.403          & 0.917          & 0.611          & 0.312          & 0.594          & 0.511          & 15.8  & 20.6                      & 125M                     & 2.7                                                                      \\
SeDR-Transformer-Head & 0.405          & 0.915          & 0.622          & 0.334          & 0.605          & 0.519          & 15.8  & 20.7                      & 132M                     & 2.7                                                                      \\
SeDR-Global-Attention & 0.406          & 0.920          & 0.600          & 0.310          & 0.584          & 0.499          & 21.2  & 26.9                      & 149M                     & 3.4                                                                      \\
SeDR-Longformer       & 0.408          & \textbf{0.922} & 0.625          & 0.315          & 0.578          & 0.518          & 88.2  & 65.2                      & 149M                     & 21.1                                                                     \\
SeDR                  & \textbf{0.409} & 0.921          & \textbf{0.632} & \textbf{0.343} & \textbf{0.607} & \textbf{0.527} & 15.8  & 20.7                      & 125M                     & 2.7                                                                      \\
\bottomrule
\end{tabular}
\caption{Comparing with different segment-interaction pattern. The compared models are derived from SeDR with different segment-interaction patterns to generate same-scale segment index (21G) for dense retrieval. Training time indicate time consumption per training epoch. MRR, NDCG and Recall are set to MRR@100, NDCG@10 and Recall@100.}
\label{res2}
\end{table*}

\subsection{Segment-Interaction Pattern}
To investigate the effectiveness of Segment-Interaction Transformer, we compare it with different segment-interaction patterns on SeDR. All these variant models use Late-Cache Negative and the same hyper-parameters setting. As shown in Table~\ref{res2}, all the segment-interaction patterns improve the performance of original SeDR-MaxP, in which segments encode independently. It demonstrates that segment-interaction can essentially alleviate information loss of document 
splitting.

Here we analyze segment-interaction models. SeDR-Transformer-Head adopts an extra Transformer layer on the segment representation, where the segment-interaction is limited to the encoder output and requests extra parameters. In contrast, SeDR-GLobal-Attention and SeDR-Longformer use global attention mechanism that allows every segment \emph{[CLS]} token to attend to whole document tokens. However, global attention mechanism request more computations and parameters. Compared with SeDR-Global-Attention, the sparse attention architecture of Longformer requests much more training time and retrieval latency. Among segment-interaction models, SeDR outperforms other models with the smallest parameters and highest efficiency.

\begin{figure}[t]
  \centering
  \includegraphics[width=1\columnwidth]{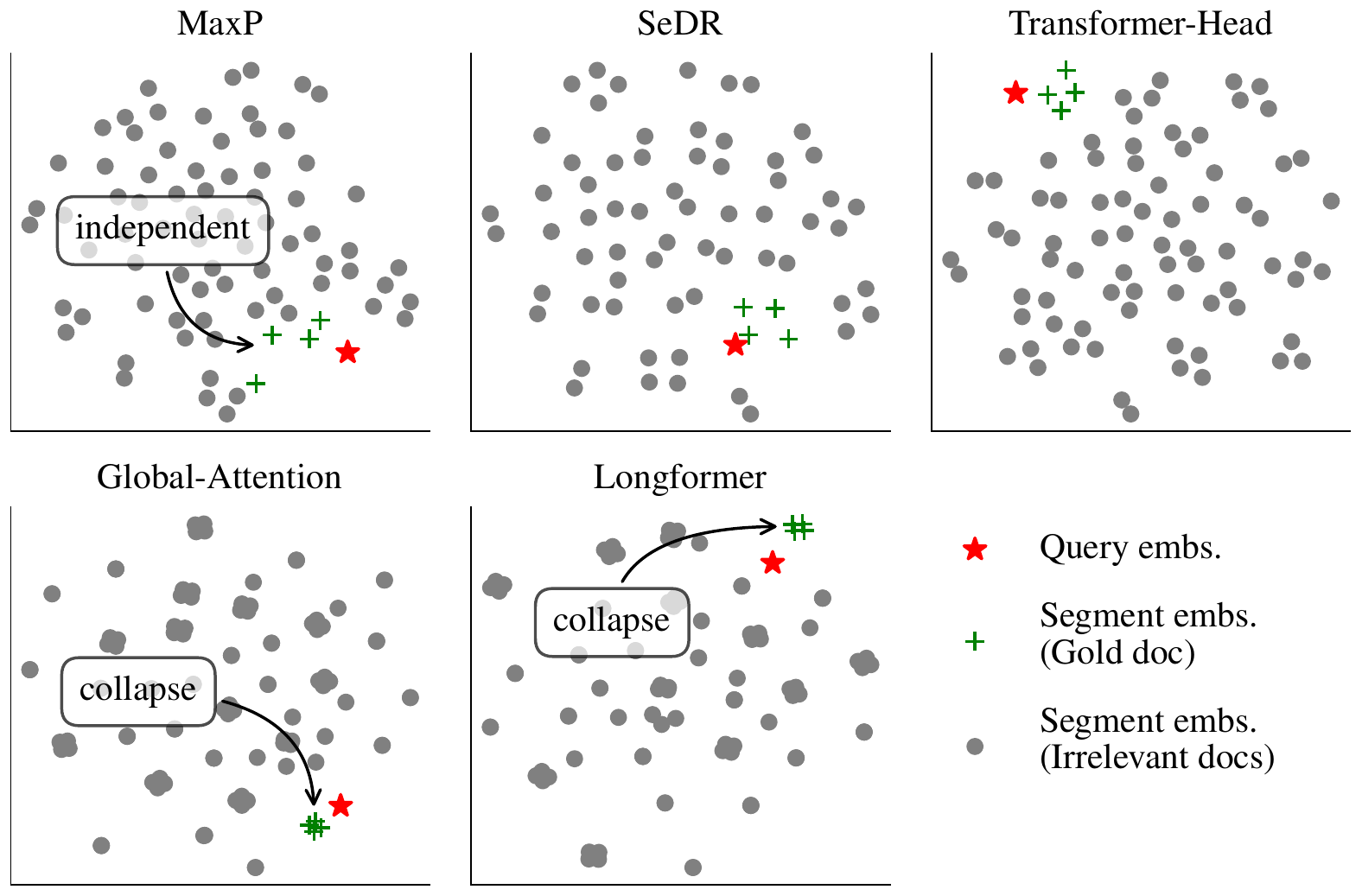}
  \caption{\label{plt1}The t-SNE plot of query and segment representations for models with different segment-interaction patterns. The query id used is 201376 from MARCO Dev.}
\end{figure}

To further investigate the upgrade of Segment-Interaction Transformer, we show the segment embeddings distribution in dense space using t-SNE in Figure~\ref{plt1}. As shown in Figure~\ref{plt1}, the segment embeddings tend to collapse into one point for those using global attention. As \citet{Multi-View} claims, the collapse of segment embedding can reduce the performance of multiple representations. Instead, the segment embeddings of SeDR-MaxP diffuse arbitrarily due to independently encoding. By contrast, SeDR and SeDR-Transformer-head learn the global-aware and segment-sensitive segment representations that scatter to a document area in Figure~\ref{plt1}. The area distribution of SeDR and SeDR-Transformer-head benefit DR, where it's consistent with the greater performance on TREC DL 19 and 20.

\subsection{Negative Setting}
In this section, we discuss the negative setting for Late-Cache Negative and hard negative sampling. Different from the in-batch negative, the number of cache negatives is not limited to GPU memory or batch-size settings. It gives the way to flexibly set the negative number. Therefore, we explore how to set the cache size $C$ of Late-Cache Negative. As Late-Cache Negative can increase training on discriminating random negatives, we further explore how to adjust the hard negative setting by controlling the top-$K$ document sampling, which we call hardness. 

\begin{figure}[t]
\centering
{ \label{fig:b}
\includegraphics[width=0.48\columnwidth]{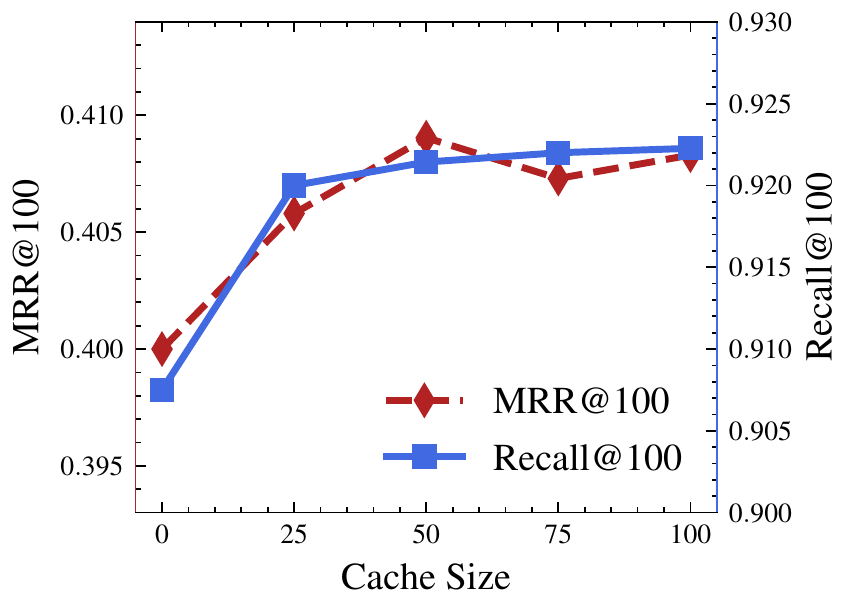}
}
{ \label{fig:a}
\includegraphics[width=0.48\columnwidth]{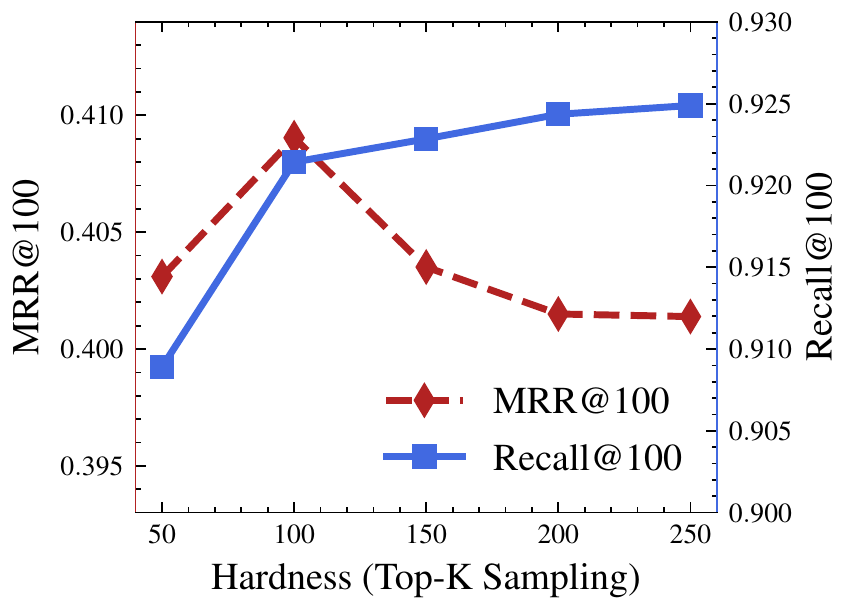}
}
\caption{\label{plt3}Evaluation of cache size $C$ and hardness $K$ on MARCO Dev. Default $C$ is 50 and default $K$ is 100.}
\label{fig}
\end{figure}

Figure~\ref{plt3} illustrates the results of different settings of cache size $C$ and hardness $K$. As for the cache size, too small cache size ($C<50$) can significantly degrade the performance. If $C>50$, larger cache size yields near performance. It also demonstrates that it's necessary to replenish the insufficient negatives for long documents training. 
In terms of hardness, we can see that less hardness (larger K) contributes to higher Recall scores, while less hardness (larger K) reduces MRR@100 scores, which are used to evaluate the top-ranking performance. Since too large hardness (small K) hurt performance, we set the $K$ as 100 and $C$ as 50 to keep a high MRR@100 score and Recall@100 score at the same time. It suggests setting high $K$ in training when applying DR to tasks that need more relevant documents recall. 

\section{Conclusion}
In this work, we propose \textit{Se}gment representation learning for long documents \textit{D}ense \textit{R}etrieval (SeDR) to tackle the long document issues in DR models. Concretely, we propose Segment-Interaction Transformer that encodes document into document-aware and segment-sensitive representations, which is verified to be superior to other segment-interaction patterns. To address the negative limitation of GPU memory, we devise Late-Cache Negative to provide additional negative instances that is shown to be necessary for long documents training. Finally, our model SeDR outperforms other DR models on MS MARCO and TREC-DL.

\bibliography{aaai23}

\end{document}